\documentclass[pre,aps,graphicx,float,draft,showpacs]{revtex4}

\usepackage[dvips,final]{graphicx}
\usepackage{bm}

\begin{document}

\preprint{}

\def\be{\begin{equation}}

\def\ee{\end{equation}}

\def\bea{\begin{eqnarray}}

\def\eea{\end{eqnarray}}

\title{ Liquid crystals in random porous media:  Disorder is stronger in low--density aerosils}

\author{D. E. Feldman and Robert A. Pelcovits}

\affiliation{Department of Physics, Brown University, Providence, RI 02912}

\date{\today}

\begin{abstract}

The nature of glass phases of liquid crystals in random porous media depends on the effective disorder strength. We study how the disorder strength depends on the density of the porous media and demonstrate that it can increase as the density decreases. We also show that the interaction of the liquid crystal with random porous media can destroy long--range order inside the pores.

\end{abstract}


\maketitle

In many natural and technological systems liquids are confined in random porous media. Two examples are water in soil and oil in porous rocks. Liquid flow in porous media involves complicated dynamical phenomena and this motivates significant research efforts. The physics becomes even richer when a porous matrix confines a liquid crystal. Interaction with random media affects not only dynamic but also static properties of liquid crystals. Long--range order in liquid crystalline states is unstable in the presence of even arbitrarily weak quenched disorder \cite{IM, Larkin} and novel glass phases emerge instead \cite{F01}.  In particular, slow glassy dynamics was reported in \cite{dyn}. Another disorder effect is the suppression of the isotropic-nematic \cite{isn} and nematic-smectic \cite{ns} phase transitions. Properties of the glass states depend on the relative strength of disorder and elastic forces. When disorder is strong, liquid crystals are expected to have a finite correlation length \cite{F01}, while in the weakly disordered case, quasi-long-range ordered glass phases can emerge \cite{F01,F00,R}. One would naively expect that the effective disorder is weaker in low density random porous media such as aerosil gels \cite{aerosil}, i.e., when the pore size is large. However, recent experiments found no signs of ordering in a low density aerosil system \cite{RICB} while liquid crystalline order was detected in a high density bulk fractal silica gels \cite{LKBL}. The results of light scattering experiments with nematics in dilute aerosils \cite{scattering} are consistent with a faster decay of correlation functions at large distances in lower-density aerosil gels.

In this paper we examine how the effective disorder strength at scales greater than the pore size depends on the pore size and show that this dependence has nontrivial character. In particular, the disorder strength can {\it grow} as the pore size increases. There are two reasons for this. First, the effective disorder strength is the ratio of the typical random and elastic energies. Since the latter grows slowly as a function of the pore size, the former can dominate for low aerogel densities. Second, there are long range correlations in the structure of random porous media and these correlations can enhance the effective disorder. 
We also consider the physics at scales smaller than the pore size. At such scales, random media cannot be viewed as uniform and the fractal structure affects the confined liquid crystal. We show that the interaction of the liquid crystal with the random pore surface can destroy long-range order in the bulk of the pore and calculate the order parameter correlation function for a simple model of random fractal porous media.

The organization of this article is as follows. First, we address the effective disorder strength in systems of continuous symmetry in porous media in the presence of random field and random anisotropy disorder, the latter being relevant for nematic liquid crystals. We also consider the ``directed random field" universality class which can be realized in random gels based on amino-acids \cite{privitecommunication}. Second, we investigate the problem of a liquid crystal interacting with a random Sierpinsky gasket surface. This is the simplest model of what might occur inside a pore. We find scaling behavior for the correlation functions in the pore. 

Most of our results are valid for a general system with a continuous symmetry order parameter. We focus on disordered nematics and the random field XY model \cite{GlD, NS}, the latter being the simplest model with a continuous symmetry group. These models represent two different types of quenched disorder: random anisotropy and random field. In nematics antiparallel orientations of the director correspond to the same physical state and hence to the same disorder energy,  i.e. the disorder is random anisotropy. 
Random field disorder is possible in ferroelectric liquid crystals where the energy is different for antiparallel dipole moments. We will show that the role of geometric correlations is stronger in the random anisotropy case.

Nematics can be described by the Hamiltonian \cite{F01}: 
\bea
\label{1}
H_n=\int d^3 r \Bigl\lbrace\frac{1}{2}[K_1(\mathbf\nabla \cdot \mathbf n(\mathbf r))^2 + K_2(\mathbf n(\mathbf r) \cdot \mathbf\nabla \times \mathbf n(\mathbf r))^2 + K_3 (\mathbf n(\mathbf r)\times\mathbf\nabla\times\mathbf n(\mathbf r))^2] 
-\sum_{\alpha,\beta=x,y,z}h_{\alpha\beta}(\mathbf r)n_\alpha(\mathbf r) n_\beta(\mathbf r)\Bigr\rbrace
\eea
where $K_i$ are the Frank moduli, the unit vector  $\mathbf n$ is the director,  and the random tensor $h_{\alpha\beta}(\mathbf r)$ describes the interaction with the porous medium. We assume that the preferred orientation of the nematic molecules near the surface of the pores is parallel to the surface.  We also assume that the nematic order parameter is relatively large in the pores, so that it is appropriate to use the director description, rather than the more general tensor nematic order parameter. 

The random field XY model has the following Hamiltonian

\be
\label{2}
H_{xy}=\int d^3 r \Bigl\lbrace\frac{J}{2} \sum_{\alpha=x,y,z}(\partial_\alpha\phi(\mathbf r))^2 
-h_1(\mathbf r)\cos\phi(\mathbf r)-h_2(\mathbf r)\sin\phi(\mathbf r)\Bigr\rbrace,
\ee
where $\phi(\mathbf r)$ is the polar angle of the spin, and $h_1(\mathbf r)$ and $h_2(\mathbf r)$ are the $x$ and $y$ components of the random field respectively, each with zero average.  Random field disorder is possible in ferroelectric liquid crystals. Since ferroelectric liquid crystals have other degrees of freedom in addition to their dipole moment, the random field XY model does not provide a full realistic description of ferroelectric liquid crystals. However, this simple random field model does provide useful insight.  The model (\ref{2}) is also relevant to discotics and smectics confined in stretched aerogels \cite{R} but only at length scales greater than the pore size.

%In the absence of random porous media,
%the discotic molecules form a hexagonal lattice. When a discotic is confined in a stretched %aerogel, the lattice is distorted. Its Hamiltonian can be expressed as a sum of decoupled %Hamiltonians $H=H_\|(u_\|)+H_\bot(u_\bot)$ for the lattice displacements $u_\|$ and $u_\bot$ in %the directions parallel and perpendicular to the stretch direction \cite{R}. The Hamiltonian %$H_\|$ has the form (\ref{2}) up to a trivial rescaling \cite{R}.

The models Eqs.~(\ref{1}) and (\ref{2}) can be used for length scales both greater and smaller than the pore size. However, the probability distribution of the random field is independent of the coordinates only in the former case. For short length scales the structure of the porous matrix must be taken into account. The effective model at length scales greater than the pore size must be derived from the short length scale model.  Such a derivation is difficult and only order--of--magnitude estimates for the model parameters have been obtained by fitting experimental data \cite{KP}. We focus on the disorder strength dependence on the aerogel density.
The disorder strength in the long length scale model determines the type of ordering at long length scales. For weak quenched disorder, quasi-long-range order is expected for nematics and the random field XY model. The ordering disappears for strong quenched disorder.

Let $a$ be the microscopic cut-off scale for the short length scale model, $\xi$ the pore size, and $d_f$ the fractal dimension of the aerogel. An Imry--Ma type estimate \cite{IM} shows that the typical elastic energy at the scale $\xi$ is $J(\xi)\sim J(a)(\xi/a)^d/(\xi/a)^2=J(a)(\xi/a)^{d-2}$, for a $d$--dimensional system and the disorder energy $h(\xi)\sim h(a)\sqrt{(\xi/a)^{d_f}}$. The effective disorder strength at the scale $\xi$ is the ratio of the typical disorder and elastic energies: 

\be
\label{3}
h_{r}(\xi)=h(\xi) J(a)/J(\xi)\sim \xi^{d_f/2-d+2}.
\ee
This effective disorder strength $h_{r}$ enters the random contribution to the Hamiltonian of the long length scale model,
if one keeps the same elastic constant as in the short length scale model.
We see that the disorder strength in the long length scale model grows as a function of the pore size for $d_f>2d-4$ (destroying long--range order in this case) while the aerogel density $\rho\sim \xi^{d_f-d}$ decreases for all fractal objects. Thus, in $d=3$, the disorder strength grows as a function of $\xi$ for all $d_f>2$, and we note that this condition is satisfied in recent experiments on highly porous gels \cite{scattering,LKBL,IGMR}. Thus, increasing the pore size can lead to an \textit{increase} in the effective disorder strength with an accompanying decrease in the aerogel density. In particular, if the disorder is strong for a smaller pore size it is also strong for a greater pore size. A simple way to understand the critical fractal dimension 
$d_f=2d-4=2$ for $d=3$ is based on the following observation: The effective disorder strength is infinite for a system of disconnected pores and the fractal dimension of the walls of disconnected pores is always 2 or greater.

The above estimate neglects correlations of the random field $h$ at different points. This approximation is valid for the random field disorder. For the nematic problem Eq. (\ref{1}), geometric correlations in random media result in correlations of the random anisotropy since nematic molecules tend to be parallel to the pore surface. This affects the effective disorder at the scale $\xi$. The existence of this effect becomes obvious when the porous matrix is not random at length scales shorter than $\xi$. Then the random 
anisotropy is the same at all points of the random medium inside the pore. Hence, $h(\xi)\sim\xi^{d_f}$ and $h_{r}\sim\xi^{d_f-1}$, where $d_f$ is the fractal dimension of the porous medium. The effective disorder strength then grows as a function of $\xi$ for all  $d_f>1$. A similar effect is possible for a random fractal. Indeed, as $d_f\rightarrow 1$ we expect that random and non--random fractals should exhibit similar properties for the case of random anisotropy disorder.  For example, consider a random fractal whose construction is based on fractional Brownian motion \cite{MvN}.  The coordinates of the $N$th segment of the random fractal are given by the equation \cite{MvN}

\be
\label{4}
\mathbf B^H(N)=\sum_{n=1}^{N} \Delta\mathbf B(n) (N-n+1)^{H-1/2},
\ee
where the Hurst exponent $H$ is related to the fractal dimension by $d_f=1/H$ \cite{MvN}, and  $\Delta\mathbf B(n)$is a random Gaussian vector with $n$ an integer parameter. A Hurst exponent of $1/2$ corresponds to ordinary Brownian motion. We imagine that this fractional Brownian walk forms the backbone of a tube--like fractal object, and the nematic molecules are anchored along the surface of the tube. To obtain a tractable model, we assume that the surface area of the $n$th segment of the tube is proportional to $ (\Delta \mathbf B^H(n))^2$, where $\Delta\mathbf B^H(n)=
\mathbf B^H(n)-\mathbf B^H(n-1)$. The average surface area of a segment is finite and hence the structure is well-defined for $H<1$.  The total strength of the random anisotropy associated with each segment $h_{\alpha\beta}(n)\sim \Delta B^H_{\alpha}\Delta B^H_{\beta}$ is proportional to its area. An estimate of the total disorder strength in a pore of linear dimension $\xi$, enclosing a fractal consisting of $n_{max}$ segments with the $z$ axis chosen to lie along the average director is then given by:
\bea
\label{5}
h(\xi)=\left[\overline{(\sum_n h_{xy}(n))^2}\right]^{1/2}\sim [\sum_{n,m}\overline{\Delta B^H_x(n)\Delta B^H_y(n) \Delta B^H_x(m) \Delta B^H_y(m)}]^{1/2}\nonumber\\
=\Bigl[\sum_{n,m}\left(\overline{\Delta B^H_x(n)\Delta B^H_x(m)}\right)^2\Bigr]^{1/2}\sim \Bigl[n_{\rm max}\sum_n (n^{2H-2})^2 \Bigr]^{1/2}\sim n_{\rm max}^\gamma \sim \xi^{\gamma d_f},
\eea
where,
\bea
\label{5a}
\gamma=\cases{1/2, & $H \leq 3/4$ \cr 2H-1=2/d_f-1, & $H \geq 3/4$ \cr} 
\eea
and the bar denotes an average over all realizations of the random fractal.
Thus, as $d_f\rightarrow 1$ (i.e., $H < 3/4$), the renormalized disorder strength scales with $\xi$ in a manner different than Eq. (\ref{3}):

\be
\label{6}
h_r(\xi)\sim \xi^{1-d_f}.
\ee
One can see that the effective disorder decreases most rapidly as the function of the pore size for $d_f = 4/3$. Even in this case $h_r\sim\xi^{-1/3}$ and hence to decrease the disorder strength by one order of magnitude one needs to increase the pore size by several orders of magnitude. At $d_f=1$, Eq. (\ref{6}) predicts no dependence of the renormalized disorder strength on the pore size.
In the presence of any preferred macroscopic orientation for the segments of the random surface of the pore, the effective random anisotropy enhancement is stronger than Eq. (\ref{6}). We do not expect the geometric correlations to play an important role in systems with random fields which can be randomly oriented along one of the two directions parallel to the segment axis.

Geometric correlations are, however, important for the ``directed random field" universality class. Let a random matrix consist of random polymer chains made of identical monomers.
We assume that the monomers are asymmetric with non--equivalent ends $A$ and $B$.
The $A$--end always links the $B$--end of the next monomer in the chain. The pore diameter equals the chain size.
Consider a ferroelectric smectic $C^*$ confined in such a random medium. Its dipole moment couples linearly to the local random field which is parallel to the network monomers \cite{footnote} and is directed from their $A$--ends to their $B$--ends. Smectics $C^*$ are chiral and form helical structures \cite{dGP}. However, the typical period of the helical structure is large and at smaller scales the liquid crystal can be considered as a uniform ferroelectric.
Then the standard Imry--Ma argument can be used. 
%This universality class can emerge for certain liquid crystals in random gels based on %amino-acids \cite{privitecommunication}.
%Let a random matrix be a rigid polymer network made of monomers of the same length. 
 
%For monomer number $n$ the random field is directed from monomer $(n-1)$ to monomer $(n+1)$.
For any fractal dimension $d_f$, the vector sum of all random fields associated with the polymer of length $\xi^{1/d_f}$ is $\mathbf h(\xi)\sim \xi$, where $\xi$ is the typical distance between branching points of the network. Hence, the disorder strength in the long length scale model with  ultraviolet cut--off $\xi$ is scale independent: $h_r(\xi)\sim h(\xi)/\xi={\rm const}$.

What happens inside a pore? We need to study the limit $\xi\rightarrow \infty$. Let us first consider the exactly solvable Larkin model \cite{Larkin} of a system with continuous symmetry:

\be
\label{7}
H_L=\int d^3 r \Bigl \lbrace \frac{1}{2}(\nabla\phi)^2-h(\mathbf r)\phi(\mathbf r)\Bigr \rbrace,
\ee
where $h(\mathbf r)$ is a random field which is nonzero on the random fractal only, and its probability distribution is characterized by  $\overline{h(\mathbf r_1)h(\mathbf r_2)}=\Delta \delta(\mathbf r_1 - \mathbf r_2)$.
We want to calculate the correlation function $G(\mathbf r_1,\mathbf r_2)=\overline{(\phi(\mathbf r_1)-\phi(\mathbf r_2))^2}$. Let $s_1$ and $s_2$ denote the distances from the points $\mathbf r_1$ and $\mathbf r_2$ to the random fractal. We assume that $R=|\mathbf r_1-\mathbf r_2|>{\rm max}(s_1,s_2)$.
At low temperatures, the equilibrium configuration of the field $\phi$ can be found by minimizing the Hamiltonian (\ref{7}):
$\phi(\mathbf r)=\int d^3 r' h(\mathbf r')/4\pi|\mathbf r-\mathbf r'|$.
Hence, 

\be
\label{8}
G\sim\Delta\int d^{d_f}r \left(\frac{1}{|\mathbf r_1-\mathbf r|}-\frac{1}{|\mathbf r_2-\mathbf r|}\right)^2,
\ee
where the integration extends over the random fractal. The result depends on the fractal dimension: for $d_f<2$, $G\sim s_1^{d_f-2}+s_2^{d_f-2}$; for $d_f=2$, $G\sim \log R/s_1+\log R/s_2$; for $d_f>2$, $G\sim R^{d_f-2}$. As $R\rightarrow\infty$, the correlation function diverges for $d_f\ge 2$. In terms of the ``physical" order parameter ${\mathbf n}=(\cos\phi,\sin\phi)$,
the correlation function $\lim_{R\rightarrow\infty}\langle\mathbf n(\mathbf r_1)\mathbf n(\mathbf r_2)\rangle=\lim_{R\rightarrow\infty}\langle\cos(\phi(\mathbf r_1)-\phi(\mathbf r_2))\rangle=\lim_{R\rightarrow\infty}\exp(-G/2)\rightarrow 0$. Hence, for such $d_f$, long range order is destroyed inside the pore.

We now consider the more realistic and complicated nonlinear model (\ref{2}). The random field XY model cannot be solved exactly and we will use a renormalization group (RG) procedure. We will develop a double $\epsilon$-expansion in $\epsilon=4-d$ and $\epsilon_f=4-2d+d_f$, where $d$ is the dimensions of space and $d_f$ the random fractal dimensions. This choice of small parameters is motivated by the fact that $d=4$ is the critical dimensions for the bulk random field XY model \cite{GlD, NS}, and the Imry--Ma argument of Eq. (\ref{3}) shows that the random fractal destroys long--range order for $d_f\ge 2d-4$ (cf. the Larkin model, where $d_f=2$ is the critical fractal dimensions for $d=3$). In contrast to the Larkin model, the results depend on the structure of the random fractal. We will limit our discussion to a simple model based on the Sierpinsky gasket \cite{fractals}. We divide the system of linear size $L_0$ into blocks of size $L_1=bL_0$ , with $b <1$. We assume that disorder is present only in a fraction $p=(L_0/L_1)^{d_f-d}$ of the blocks. Then we divide each block of size $L_1$ into blocks of size $L_2=bL_1$. For each block of size $L_1$ that contains segments of the random fractal we randomly choose a fraction $p$ of subblocks of size $L_2$, such that only these subblocks contain segments of the random fractal. We iterate this procedure and generate a random fractal of dimension $d_f$. We implement the Wilson shell renormalization group (RG) by changing the cut-off scale from $L_n$  to $L_{n-1}$ at each step, to generate the one--loop RG equations. For the bulk random field XY model this problem was solved in Ref. \cite{GlD1}. In the case when disorder is present on a two--dimensional surface only, the problem was studied in Ref. \cite{F02}. We will thus only briefly discuss the RG procedure which is similar to Refs. \cite{GlD1, NS, F01, F02}. 

Due to the periodicity of the Hamiltonian there is no renormalization of the order parameter.
There also is no renormalization \cite{GlD1} of the elastic term in (\ref{2}). Hence, the scaling dimension of the temperature \cite{GlD1} is $2-\epsilon$. As for disorder,
we know that all random contributions of the form $h^{(k)}_1\cos k\phi + h^{(k)}_2\sin k\phi$
are relevant operators in the vicinity of the zero-temperature fixed point \cite{GlD, NS}. After  replica averaging this corresponds to the following structure of the relevant terms in the Hamiltonian:

\be
\label{9}
H_R=\int d^d r [\sum_a\frac{(\nabla\phi_a(\mathbf r))^2}{2T}-\sum_{ab}P(\mathbf r)\frac{F(\phi_a(\mathbf r)-\phi_b(\mathbf r))}{2T^2}],
\ee
where $a,b$ are replica indices, $F$ is the function describing disorder at the ultra-violet cut-off scale $L_n$, $P(\mathbf r)=1$ in the subblocks of size $L_n$ containing elements of the random fractal, otherwise $P(\mathbf r)=0$. Eq. (\ref{9}) is obtained by averaging with respect to disorder inside the subblocks of size $L_n$. We can average over the disorder distribution inside the blocks of size $L_{n-1}$ instead. This will include averaging with respect to the distribution of 
$P(\mathbf r)$ which is nonzero with probability $p$ in different points of the blocks of size $L_{n-1}$ which contain segments of our fractal. Instead of $\exp(\int d^d r\sum_{ab}P(\mathbf r){F(\phi_a(\mathbf r)-\phi_b(\mathbf r))}/{2T^2})$ in the replica action we will get $\Pi_{\bf r}[(1-p)+p\exp(P'(\mathbf r)\sum_{ab}{F(\phi_a(\mathbf r)-\phi_b(\mathbf r))}/{2T^2})]=\exp(\int d^d r P'(\mathbf r)[\sum_{ab} p{F(\phi_a(\mathbf r)-\phi_b(\mathbf r))}/{2T^2}+\sum_{abcd}(p-p^2)
F(\phi_a(\mathbf r)-\phi_b(\mathbf r))F(\phi_c(\mathbf r)-\phi_d(\mathbf r))/8T^4+...])$, where $P'(\mathbf r)=1$ if $\mathbf r$ belongs to a block of size $L_{n-1}$ containing segments of the fractal. We need to keep only the first term $\sum_{ab} p{F(\phi_a(\mathbf r)-\phi_b(\mathbf r))}/{2T^2}$ in the integral in order to derive the one--loop RG equation. Indeed, we will see that $F\sim\epsilon_f$ at the fixed point. We note that $(p^2-p)\rightarrow 0$ as $\epsilon,\epsilon_f\rightarrow 0$. Hence, the second term in the square brackets is $O(\epsilon^3,\epsilon_f^3)$ and will not contribute to the one--loop RG equation. The same is true for the higher-order terms denoted by the dots. Thus, one can directly repeat the standard derivation of the RG equations for the bulk random field XY model. The only modification is related to the factor $p$ multiplying $F$. This factor will change the term linear in $F$ in the one--loop RG equation. We obtain:

\be
\label{10}
dF(\phi)/d\ln L = \epsilon_f F(\phi) + F''(\phi)^2/2 - F''(\phi)F''(0),
\ee
where the primes refer to derivatives of $F$ with respect to its argument and the factor $1/8\pi^2$ has been absorbed by $F$. The fixed point solution of the above equation is known \cite{F01,GlD1}. 
Due to the symmetry of the problem we need a periodic solution with period $2\pi$. At the interval $0<\phi<2\pi$ it is given by the equation $F(\phi)=2\pi^4\epsilon_f [1/36-(\phi/2\pi)^2(1-\phi/2\pi)^2]/9$. This solution corresponds to $\epsilon_f>0$.  Otherwise $F\rightarrow 0$.
The temperature goes to zero at the fixed point.  We can now calculate the correlation function $G=\overline{(\phi(\mathbf r_1)-\phi(\mathbf r_2))^2}=2\int d^d q \overline{|\phi^2(\mathbf q)|}(1-\cos\mathbf q \cdot(\mathbf r_1-\mathbf r_2))/(2\pi)^d$. At each step of the RG procedure it is sufficient to use the quadratic in $\phi$ part of the renormalized replica Hamiltonian to calculate the contribution corresponding to $1/L_n>q>1/L_{n-1}$. If both points $\mathbf r_1$ and $\mathbf r_2$ are at a distance $s<|\mathbf r_1-\mathbf r_2|$ from the random fractal,
we obtain $G=({2\epsilon_f \pi^2}/{9})\ln({|\mathbf r_1-\mathbf r_2|}/{s})$. The correlation function diverges for large $|\mathbf r_1 - \mathbf r_2|$, if $\epsilon_f>0$. Hence, the random fractal destroys long range order in the pore for $d_f>2d-4$, in agreement with our Imry--Ma estimate above. The above result for the correlation function is valid only if the bare disorder at the microscopic cut-off scale $a$ is weak since we did a perturbative calculation. 

We have calculated how the effective disorder in models, Eqs. (\ref{1}) and (\ref{2}) depends on the pore size. If the disorder is weak in the long length scale model with cutoff length $\xi$ then we expect quasi--long range order \cite{GlD, NS, F01} in both models. If the disorder is strong then only short range order is possible \cite{GlD, NS, F01}.  At strong disorder the perturbative RG analysis is no longer valid and quasi--long--range order is impossible. The condition of weak disorder means two things: (1) the effective disorder at the pore size scale $\xi$ is weak; (2) the renormalized disorder is weak at long length scales, i.e. $\epsilon=4-d$ is small. The critical value
of the disorder strength is non--universal and depends on microscopic details.

In conclusion, we have found the dependence of the effective disorder in random fractal media on the pore size and the fractal dimensions. The dependence is non--trivial and the effective disorder can grow as the density of the porous media decreases. This suggests that instead of changing the pore size a more effective way to control the disorder strength is to use mixtures of nematics with other substances whose molecules could ``shield" the nematic from the random surface. This would decrease the liquid crystal interaction with the porous media. 
We also showed that at certain conditions the porous media can destroy long range order inside a pore. 

We thank Richard Stratt for a useful discussion. This work was supported by the National Science Foundation under grant DMR--0131573.


\begin{thebibliography} {150}
\bibitem{footnote} Higher-order couplings can also be present.
\end{thebibliography}
\end{document}